# Breaking Giant Chains: Early-stage Instabilities in Long-period Giant Planet Systems

Vighnesh Nagpal[1], Max Goldberg[2], and Konstantin Batygin[3]
[1] Departments of Astronomy & Physics, University of California, Berkeley, CA 94720, USA
[2] Department of Astronomy, California Institute of Technology, 1200 E. California Blvd., Pasadena, CA 91125, USA
[3] Division of Geological and Planetary Sciences, California Institute of Technology, 1200 E. California Blvd., Pasadena, CA 91125, USA

## Abstract

Orbital evolution is a critical process that sculpts planetary systems, particularly during their early stages where planet–disk interactions are expected to lead to the formation of resonant chains. Despite the theoretically expected prominence of such configurations, they are scarcely observed among long-period giant exoplanets. This disparity suggests an evolutionary sequence wherein giant planet systems originate in compact multiresonant configurations, but subsequently become unstable, eventually relaxing to wider orbits—a phenomenon mirrored in our own solar system's early history. In this work, we present a suite of $N$-body simulations that model the instability-driven evolution of giant planet systems, originating from resonant initial conditions, through phases of disk dispersal and beyond. By comparing the period ratio and normalized angular momentum distributions of our synthetic aggregate of systems with the observational census of long-period Jovian planets, we derive constraints on the expected rate of orbital migration, the efficiency of gas-driven eccentricity damping, and typical initial multiplicity. Our findings reveal a distinct inclination toward densely packed initial conditions, weak damping, and high giant planet multiplicities. Furthermore, our models indicate that resonant chain origins do not facilitate the formation of Hot Jupiters via the coplanar high-eccentricity pathway at rates high enough to explain their observed prevalence.

*Unified Astronomy Thesaurus concepts:* Exoplanet dynamics (490); Exoplanet astronomy (486); Exoplanet migration (2205); Exoplanet formation (492); Dynamical evolution (421); Planetary dynamics (2173); Planetary science (1255)

## 1. Introduction

Exoplanetary systems exhibit stunning architectural diversity, a phenomenon that is almost certainly sculpted by orbital evolution that occurs concurrently with, and after, formation. In particular, interactions with a gaseous circumstellar disk (e.g., Goldreich & Tremaine 1980; Ward 1997), scattering events with a planetesimal swarm (Fernandez & Ip 1984; Malhotra 1993; Tsiganis et al. 2005), and tidal dissipation (e.g., Goldreich & Soter 1966; Yoder & Peale 1981) are all dissipative mechanisms that are known to drive orbital migration. In the presence of convergent migration, it is possible for pairs of planets to capture into *mean motion resonance* (MMR), an orbital configuration characterized by the libration of resonant angles and associated with planetary period ratios near (or at) exact commensurability (Henrard & Lemaitre 1986). In the case where there are more than two planets, this mechanism can also lead to the formation of *resonant chains*, configurations in which adjacent pairs of planets are in MMR. In fact, numerous resonant chains have been discovered among the growing exoplanet system sample (e.g., Shallue & Vanderburg 2018; Leleu et al. 2021; Dai et al. 2023).

Turning our attention homeward, orbital migration and mean motion resonances were likely key pieces of the puzzle that is the assembly of the solar system. Within the context of the Nice model (Gomes et al. 2005; Morbidelli et al. 2005; Tsiganis et al. 2005), the resonant capture of Jupiter and Saturn critically influences the subsequent dynamical evolution of the solar system (Batygin & Brown 2010; Nesvorný & Morbidelli 2012). Given the physical basis for orbital migration and the subsequent resonant capture, as well as the success of models based on this paradigm at explaining many features of our own solar system, it comes as somewhat of a surprise that resonant (or near-resonant) planets together form only a small minority of the known exoplanet population (Fabrycky et al. 2014).

To reconcile the relative lack of observed resonant systems with the high prevalence of resonances expected from migration, scenarios such as the "breaking-the-chains" framework (Izidoro et al. 2017) have been proposed. In this picture, planets are routinely captured into resonant chains early in a system's lifetime but are then broken out of resonance by the onset of instabilities—often associated with the protoplanetary disk's dispersal or some other event. For planets with radii between 1 and 4 $R_\oplus$ (so-called "super-Earths" or "mini-Neptunes"), this scenario can reproduce the corresponding period ratio distribution from the Kepler sample if a large fraction (∼90%) of resonant chains experience instability (Izidoro et al. 2021). Subsequent efforts, such as those of Goldberg & Batygin (2022), have shown that by invoking the onset of instabilities of resonant chains it is possible to reproduce the preference for not only the period ratio distribution but also the intrasystem uniformity (the so-called "peas-in-a-pod" pattern; Weiss et al. 2018) exhibited by the observational sample of super-Earths.

How giant planets fit into this picture is, as of now, less clear. First off, the migration of giant planets is expected to exhibit qualitative differences relative to their lower mass counterparts, since giant planet embryos can be massive enough to carve gaps in the protoplanetary disk in which they are embedded (Crida et al. 2006). While numerous studies using detailed hydrodynamic simulations have been dedicated to the effort of understanding giant planet migration, fundamental properties, such as its characteristic timescale and the strength of eccentricity damping, remain an active area of research (e.g., Kanagawa et al. 2018; Griveaud et al. 2023). Further, the resonant dynamics of giant planets also exhibit significant differences relative to low-mass planets (see scalings in Batygin 2015 and Deck & Batygin 2015). Moving to later epochs, there is considerable evidence suggesting that planet–planet scattering plays an important role in sculpting the observed giant exoplanet population (e.g., Rasio & Ford 1996; Ford & Rasio 2008; Jurić & Tremaine 2008).

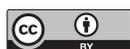

In recent years, progress on the observational front from the realm of direct imaging (a technique most sensitive to detecting young giant planets) has begun to provide a view into the early-stage orbital configurations of the most massive planetary systems. Curiously, two of the best-studied directly imaged systems with multiple detected planets, HR 8799 and PDS70, exhibit inferred orbital configurations consistent with mean motion resonance (Wang et al. 2018; Bae et al. 2019). Probing even earlier epochs, programs using the Atacama Large Millimeter/submillimeter Array (ALMA) are mapping features of circumstellar disks that potentially encode information about the early-stage masses and locations of giant planets (Andrews et al. 2018; Ohashi et al. 2023). Complementary to these studies of young systems is the stream of discoveries emerging from long-baseline radial velocity surveys targeting mature stars and their planetary systems. In particular, the California Legacy Survey (CLS) released a large sample of planets discovered and characterized through radial velocity monitoring spanning 30 yr (Fulton et al. 2021; Rosenthal et al. 2021). The giant planets contained within this catalog exhibit a wide diversity in orbital configurations, which—given the highly ordered initial configurations predicted by the framework of convergent migration—begs the question: *can the evolved giant planet population arise from the dynamical evolution of resonant chain architectures?*

In this work, we bridge this gap by using numerical simulations to study the formation and evolution of synthetic giant planet resonant chains across the phases of migration, disk dispersal, and long-term gas-free evolution. Then, by comparing our results to constraints from observations, we aim to place reasonably model-independent constraints on quantities, such as initial planetary multiplicity, migration timescale, and typical strengths of damping in giant planet systems. The rest of this paper is structured as follows. We begin by describing our simulation in Section 2. In Section 3, we study the evolution of system multiplicities and period ratio distributions for our synthetic systems, and the level of dynamical excitation in Section 4. Finally, in Section 5, we estimate upper limits on the production of Hot Jupiters through planet–planet scattering following the disruption of resonant chains.

## 2. Numerical Simulations

While details of giant planet accretion remain an active area of research (e.g., Bae et al. 2019; Choksi et al. 2023; Li et al. 2023), the mean motion resonant chains we focus on in this work are likely to be established after any planet–planet impacts phase that may occur (e.g., Frelikh et al. 2019; Ginzburg & Chiang 2020) during the protoplanetary disk phase and have reached nearly their final mass. With this motivation, we restrict our attention to systems that have already finished forming multiple giant planets but remain embedded in the protoplanetary disk. To study the assembly of resonant chains and the subsequent evolution of orbital architectures in such systems, we ran a suite of $N$-body simulations that capture

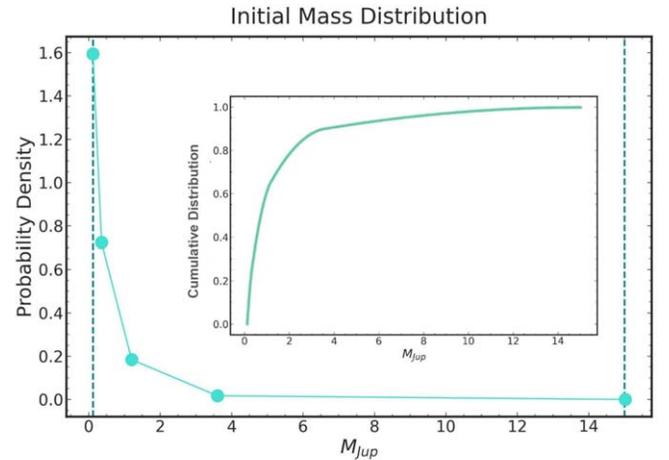

**Figure 1.** Initial distribution of masses for the planets in our synthetic systems. This pdf was constructed by first converting the $M_{\sin i}$ occurrence rate measurements by Fulton et al. (2021) into probability densities and then correcting for the inclination to arrive at a distribution for $M$. The corresponding CDF is shown in the inset. We impose lower and upper bounds of 0.12 and 15 $M_{\rm Jup}$, shown in teal dashed lines.

planet–disk interactions in a parameterized manner. With the aim of exploring their impact on the formation and evolution of giant planet resonant chains, we varied three parameters:

1. $N_{p,i}$: The initial planetary multiplicity of a system. We test initial conditions, where $N_{p,i} = 2$, 3, or 4.
2. $t_m$: Migration timescale of the inner planet. We test $t_m = 10^4$, $10^5$, and $10^6$ yr.
3. $K$: Defined through the eccentricity-damping timescale $t_e = t_m/K$. We test $K = 100$, 10, and 1.

In this work, we test every combination of these three parameters. To run our simulations, we used REBOUND's (Rein & Liu 2012) implementation of the hybrid-symplectic MER-CURIUS integrator (Rein et al. 2019) to set up and run our dynamical simulations. Orbital migration and eccentricity damping were implemented using RebeundX's (Tamayo et al. 2020) modify_orbits_forces module, which applies physical forces that when orbit-averaged, yield exponential growth or decay in semimajor axis and eccentricity (Papaloizou & Larwood 2000; Kostov et al. 2016). The mass distribution for planets in our simulations is derived from Fulton et al. (2021), where the authors report planet occurrence rates between 1 and 5 au with respect to $M \sin i$. After adjusting for bin sizes and averaged geometric effects, we use these occurrence rates to infer a probability density function in mass for the CLS planets. In order to remain anchored in the giant planet regime, we impose cutoffs at 0.12 and 15 $M_{\rm Jup}$, respectively.[4] The mass probability density function (pdf) and corresponding cumulative distribution function (CDF) thus obtained are shown in Figure 1. By randomly sampling from this mass distribution, we create 100 unique orderings $\{m_j\}$ for each number of initial planets: $N_{p,i} = 2$, 3, or 4. Given a mass ordering $\{m_j\}$, we create two synthetic systems in which the planets are initialized with spacings at constant period ratios of 2.7 and 3.3—numbers chosen in order for the planets to start wider than the 2:1 and 3:1

---

[4] The lower cutoff of 0.12 $M_{\rm Jup}$ was chosen on the basis of expectations for the minimum mass at which a planet is expected to begin to carve a gap in its disk. On the other end, the upper cutoff is approximately where the transition from giant planets to brown dwarfs is thought to occur.

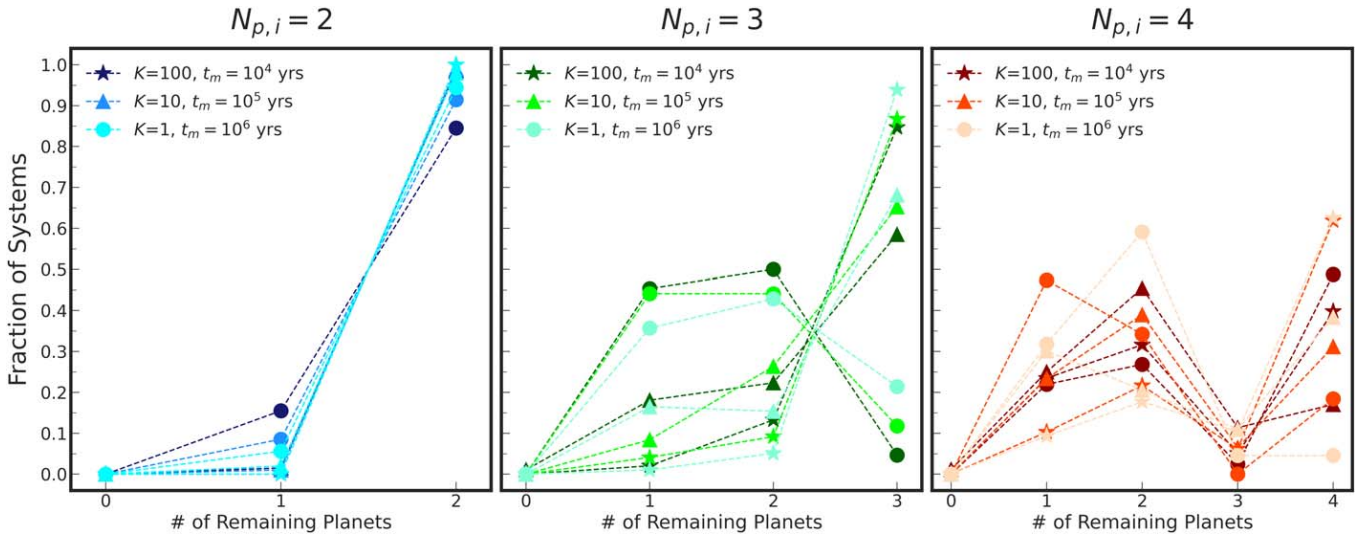

**Figure 2.** Final multiplicity fractions for simulated systems with two-, three-, and four-planet initial conditions (left to right). On each panel, the colors (from darkest to lightest) correspond to $t_m = 10^4$, $10^5$, and $10^6$ yr respectively, while stars, triangles, and circles correspond to subsamples with $K = 100$, 10, and 1.

MMR, respectively. In each case, the innermost planet is placed at $a = 1$ au, and all planets are started off on circular, coplanar orbits. Mutual inclinations only appear as second-order terms in the Hamiltonian governing the resonant dynamics of a planet pair (Murray & Dermott 1999; Batygin 2015). This motivates our expectation that small mutual inclinations between planets are likely to only have a weak effect on a system's dynamical evolution. This, combined with our desire to constrain the excitation that may emerge out of purely 2D dynamics, motivates our decision to study the coplanar resonant chain.

Cumulatively, the setup we describe above yields a starting set of 1800 unique systems for each choice of $N_i$, and starting subsets of 300 systems for each combination of $t_m$ and $K$ we test. For the rest of this study, we shall refer to subsamples within our simulations through names reflecting their unique combination of parameters. As an example, under this scheme, `N3T4K100` refers to the subset of systems with $N_{p,i} = 3$, $t_m = 10^4$ yr, and $K = 100$.

From this point on, our dynamical simulations proceed in three stages. First, each system is integrated with migrationary and damping forces for a duration of $3t_m$ yr, with a migration prescription given by

$$\frac{a}{\dot{a}} = \frac{2 t_m}{2 + \log_{10}\left(\frac{a}{\mathrm{au}}\right)}, \quad (1)$$

where the purpose of the additive logarithmic term in the denominator is to ensure convergent migration[5] to facilitate the formation of resonant chains. In practice, each system in our simulations forms resonant chains. Throughout this stage, the semimajor axes of the system are rescaled at each timestep in order to keep the innermost planet at 1 au, even as the planets all continue migrating. This is done for definitiveness, and as Newtonian gravity is scale-free (in the limit of point particles), we do not expect this to affect the results of our simulations.[6]

Throughout this, and subsequent stages of our simulation, we use a timestep of 0.016 yr.

In the next stage, we exponentially remove the migration and damping forces on a timescale of $t_d = 10^5$ yr, corresponding to the dispersal of a protoplanetary disk. The system is then integrated with these decaying disk forces for $20 t_d$ (2 Myr), and we once again continuously rescale the semimajor axes such that the innermost planet remains at 1 au. At the end of this dispersal stage, we set migration and damping timescales to infinity, which represents the complete dispersal of the protoplanetary disk and a transition to gas-free dynamical evolution.

Finally, we begin the last stage of the simulation by randomizing the orbital phases in order to trigger instabilities of various strengths by compromising the resonant locks between planets.[7] Each system is then integrated (under pure gravity and without renormalization) forward in time for a duration, which is determined by $N_{p,i}$. Systems with $N_{p,i} = 2$ are run for 20 Myr, those with $N_{p,i} = 3$ for 50 Myr, and those with $N_{p,i} = 4$ for 100 Myr. We ran an additional smaller set of simulations with longer integration times and found no significant differences in the statistical properties between the two ensembles of simulated systems. Our simulations do not allow for collisions, as they are expected to be rare relative to scattering beyond $\sim 1$ au.

### 3. Multiplicities and Period Ratios

In Figure 2, we show the distribution of final multiplicities for our simulated systems. Interestingly, systems with two-planet initial conditions rarely experience ejections: even the `N2T4K1` subsample (the most extreme combination of parameters) has an ejection fraction of only 10%. This picture changes substantially for systems with $N_{p,i} = 3$. Here, we observe that the frequency of ejections is a strong function of both $t_m$ and $K$, given a fixed value for $N_{p,i}$. For example, while 91% of systems in the N3T4K1 subsample (for which the combination of rapid migration and weak damping might lead us to expect increased dynamical excitation) experience ejections, the ejection frequency drops to only 5% for the N3T6K100 subsample, which corresponds to the combination of $t_m$ and $K$, which leads to the fewest instabilities. Overall, the results for three-planet initial conditions show that the ejection frequency scales inversely with both $t_m$ and $K$, a trend that mostly extends to the $N_{p,i} = 4$ case, accompanied by a slight broadband increase in ejection rates. However, an intriguing deviation from this pattern is that the slow migration N4T6K1 subsample exhibits the highest ejection rate, and not its rapid migration counterpart N4T4K1.

---

[5] See the Appendix in Goldberg & Batygin (2022) for details of how this prescription ensures convergent migration.
[6] There is one prominent exception: the estimation of Hot Jupiter production. To counter this issue, we employ a data-driven rescaling scheme—which we elaborate on in Section 5.

[7] Though such a randomization of phases is unphysical, our aim is to characterize the orbital architectures of systems that have *already undergone* dynamical instabilities. Possible sources for such instabilities after disk dispersal include planetary mutual inclinations, interactions with binary companions, the close passage of nearby stars, and the influence of leftover planetesimals. This is in addition to the litany of effects that may arise within a protoplanetary disk and during its dispersal. Only a very small number of simulated resonant chains were preserved after our attempt to trigger instabilities. Such systems could be viewed as analogs to the resonant chains observed in mature systems such as GJ 876 (Lee & Peale 2002).

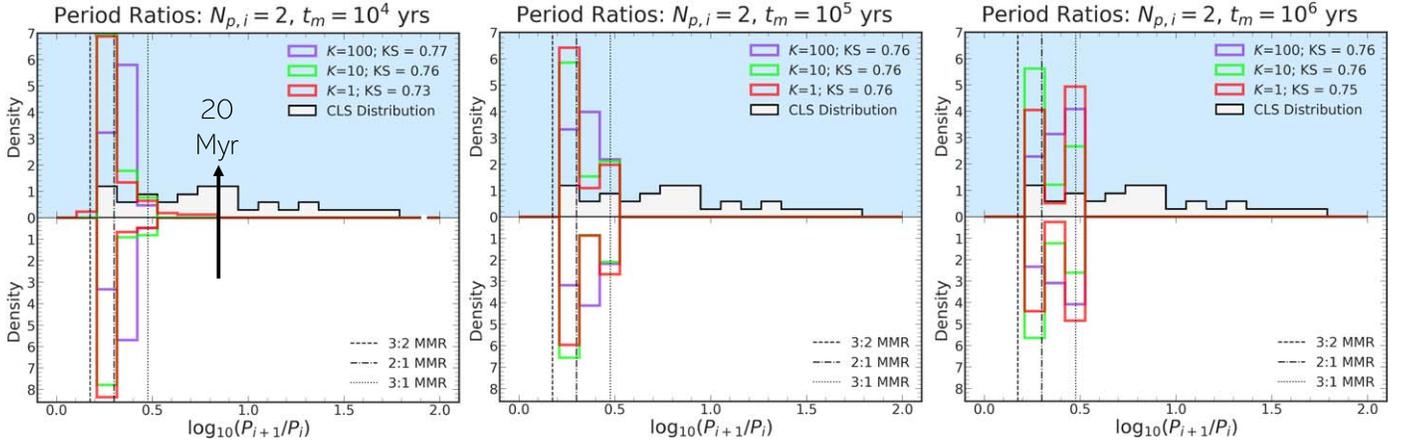

**Figure 3.** Evolution of period ratio distributions for simulated systems with two-planet initial conditions, split up by the migration timescale $t_m$ (from left to right) and damping parameter $K$ (histogram color). On each panel, we show the initial (predisk dispersal) and final (after disk dispersal and a further 20 Myr of gas-free evolution) distributions on the bottom and top subpanels, respectively. The bins are spaced uniformly (in logarithmic space) between 1 and 10. The observed CLS systems are shown in black, while the results of Kolmogorov–Smirnov tests comparing simulated systems' final architectures with the CLS are quoted on the top panel.

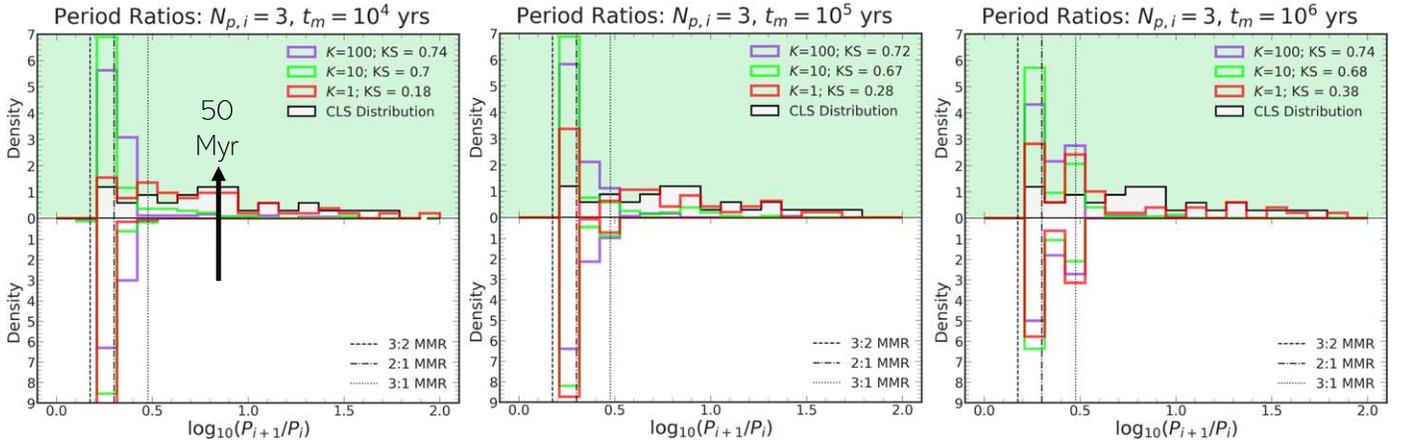

**Figure 4.** Same as Figure 3, but for simulated systems with three-planet initial conditions. In this case, after the removal of disk forces, the systems are integrated for a further 50 Myr in dissipation-free conditions.

Next, we investigate how the evolution of architectures assembled from convergent migration and (possibly disrupted through) long-term gas-free dynamical evolution manifests in period ratio space. In Figures 3–5, we show the initial (predisk dispersal) and final period ratio distributions for synthetic systems with two-, three-, and four-planet initial conditions.

For reference, we also plot the observational period ratio distribution from the CLS sample and report values of the Kolmogorov–Smirnov (KS) statistic calculated by conducting two-sample KS tests comparing the synthetic and CLS period ratio distributions. Since the KS test measures the maximum distance between the empirical CDFs of the two samples being compared, smaller values of the KS statistic indicate increasingly similar samples. The KS statistic values are also reported in Table 1.

Focusing on the case of two-planet initial conditions illustrated in Figure 3, we observe that capture into the 2:1 MMR is strongly preferred across the values of $t_m$ and $K$ tested. In addition, we also see that the number of systems caught into the 3:1 MMR rises with $t_m$, a trend which is in agreement with analytic predictions (Batygin 2015; Batygin & Petit 2023). Intriguingly, all subsamples of our synthetic systems with $N_{p,i} = 2$ appear to experience too few instabilities to match the period ratio distribution of the observed sample. Moving onto the results for three-planet initial conditions (as shown in Figure 4), we see that the situation changes somewhat. Here, the period ratios for subsamples with weak damping ($K = 1$) exhibit an appreciable level of spreading relative to their distributions during the resonant chain stage. Further, there

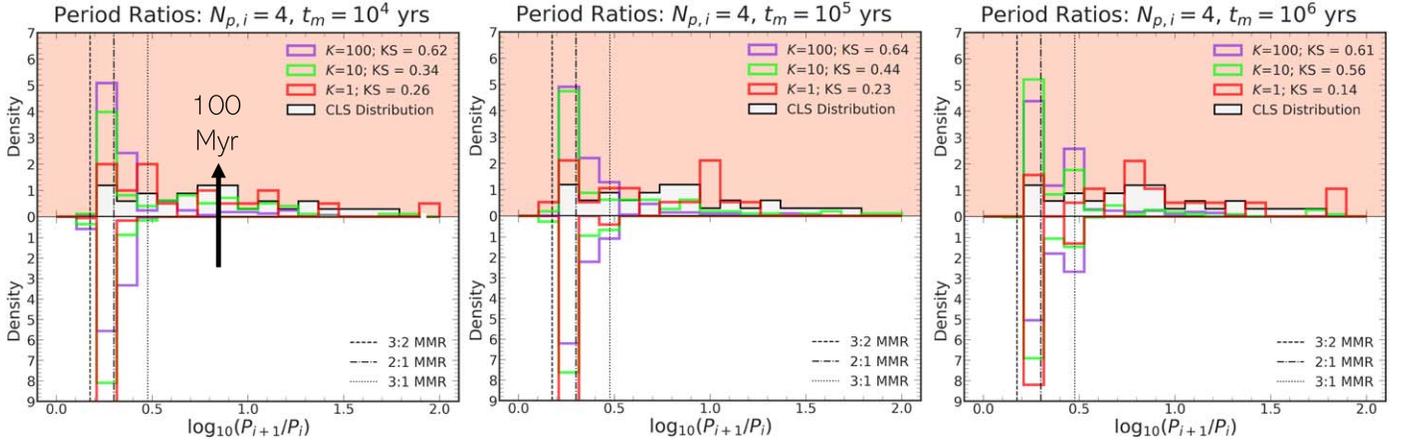

**Figure 5.** Same as Figure 3, but for simulated systems with four-planet initial conditions. In this case, after the removal of disk forces, the systems are integrated for a further 100 Myr in dissipation-free conditions.

**Table 1**
Summary of Main Statistical Results from This Study, Split up by Subsample

| $N_{p,i}$ | $t_m$ | $K$ | Chains Surviving Migration | $KS_{P_{i+1}/P_i}$ | $KS_{NAMD}$ | $\mathcal{S}$ | Fraction of Hot Jupiter–producing Systems |
|---|---|---|---|---|---|---|---|
| 2 | $10^4$ | 100 | 100/100 | 0.77 | 0.91 | 1.19 | 0% ± 0% |
| 2 | $10^4$ | 10 | 100/100 | 0.76 | 0.83 | 1.13 | 0% ± 0% |
| 2 | $10^4$ | 1 | 100/100 | 0.73 | 0.56 | 0.92 | 0% ± 0% |
| 2 | $10^5$ | 100 | 100/100 | 0.76 | 0.83 | 1.13 | 0% ± 0% |
| 2 | $10^5$ | 10 | 100/100 | 0.76 | 0.82 | 1.12 | 0% ± 0% |
| 2 | $10^5$ | 1 | 100/100 | 0.76 | 0.61 | 0.97 | 0% ± 0% |
| 2 | $10^6$ | 100 | 100/100 | 0.76 | 0.86 | 1.15 | 0% ± 0% |
| 2 | $10^6$ | 10 | 100/100 | 0.76 | 0.81 | 1.11 | 0% ± 0% |
| 2 | $10^6$ | 1 | 100/100 | 0.75 | 0.62 | 0.91 | 0% ± 0% |
| 3 | $10^4$ | 100 | 98/100 | 0.74 | 0.75 | 1.05 | 0.3% ± 0.3% |
| 3 | $10^4$ | 10 | 94/100 | 0.7 | 0.48 | 0.85 | 0.2% ± 0.2% |
| 3 | $10^4$ | 1 | 86/100 | 0.19 | 0.17 | 0.25 | 3.0% ± 1.8% |
| 3 | $10^5$ | 100 | 98/100 | 0.73 | 0.79 | 1.08 | 0.4% ± 0.4% |
| 3 | $10^5$ | 10 | 95/100 | 0.67 | 0.50 | 0.84 | 0.4% ± 0.4% |
| 3 | $10^5$ | 1 | 68/100 | 0.25 | 0.23 | 0.34 | 1.7% ± 1.6% |
| 3 | $10^6$ | 100 | 98/100 | 0.74 | 0.82 | 1.10 | 0% ± 0% |
| 3 | $10^6$ | 10 | 91/100 | 0.68 | 0.56 | 0.88 | 0.3% ± 0.3% |
| 3 | $10^6$ | 1 | 57/100 | 0.37 | 0.28 | 0.46 | 4.0% ± 2.5% |
| 4 | $10^4$ | 100 | 98/100 | 0.62 | 0.4 | 0.74 | 1.0% ± 0.9% |
| 4 | $10^4$ | 10 | 88/100 | 0.32 | 0.14 | 0.35 | 1.9% ± 1.2% |
| 4 | $10^4$ | 1 | 51/100 | 0.35 | 0.22 | 0.41 | 2.2% ± 2.1% |
| 4 | $10^5$ | 100 | 97/100 | 0.64 | 0.49 | 0.81 | 0.4% ± 0.4% |
| 4 | $10^5$ | 10 | 81/100 | 0.45 | 0.20 | 0.49 | 2.1% ± 1.4% |
| 4 | $10^5$ | 1 | 32/100 | 0.24 | 0.27 | 0.36 | 1.8% ± 1.8% |
| 4 | $10^6$ | 100 | 96/100 | 0.62 | 0.51 | 0.80 | 0.3% ± 0.3% |
| 4 | $10^6$ | 10 | 78/100 | 0.56 | 0.30 | 0.64 | 0.9% ± 0.9% |
| 4 | $10^6$ | 1 | 22/100 | 0.18 | 0.27 | 0.32 | 3.7% ± 3.7% |

**Note.** From left to right, the columns are: the initial multiplicity ($N_{p,i}$), migration timescale ($t_m$) in yr, strength of damping ($K$), the number of systems in which formed resonant chains survive the first stage of the simulation, the KS statistics resulting from the comparison of subsample's period ratio ($KS_{P_{i+1}/P_i}$) and NAMD ($KS_{NAMD}$) distributions with those from Rosenthal et al. (2021), $\mathcal{S}$ (defined as $KS_{NAMD}$ and $KS_{P_{i+1}/P_i}$ added in quadrature), and the estimated Hot Jupiter production rate.

appears to be a slight preference for shorter migration timescales: of all subsamples with $N_{p,i} = 3$, N3T4K1 shows the closest agreement with the data (KS = 0.19), followed by N3T5K1 (KS = 0.25) and N3T6K1 (KS = 0.37), respectively. Broadly speaking, however, three-planet initial conditions still appear to exhibit more clustered period ratio distributions than the observations.

The preference for weak damping persists for systems with four-planet initial conditions, as illustrated by Figure 5. More curious is the fact that *slower* migration timescales are preferred, the reverse of the trend observed for subsamples with three-planet initial conditions. Indeed, of all four-planet subsamples, N4T6K1 (KS = 0.18) performs the best, followed by N4T5K1, N4T4K10, and N4T4K1.

## 4. Normalized Angular Momentum Deficit and Eccentricity Distributions

The period ratio provides one important metric for evaluating our results against the observations. However, it does not quantify the dynamical temperature of planetary systems. For this purpose, we turn to the normalized angular momentum deficit (Chambers 2001; Turrini et al. 2020). This quantity (which we shall henceforth refer to as the NAMD) is defined as

$$\mathrm{NAMD} = \frac{\sum_j m_j \sqrt{a_j}(1 - \cos(i_j)\sqrt{1-e_j^2})}{\sum_j m_j \sqrt{a_j}}, \quad (2)$$

and provides a measure of the level of dynamical excitation in multiplanet systems with different architectures. We compute the NAMD for each system in both the simulated and observational[8] samples, and show the resulting subsample-level NAMD distributions in Figure 6. As in Section 3, we perform two-sample KS tests to measure similarity and report the calculated values of the KS statistic in Figure 6 and in Table 1.

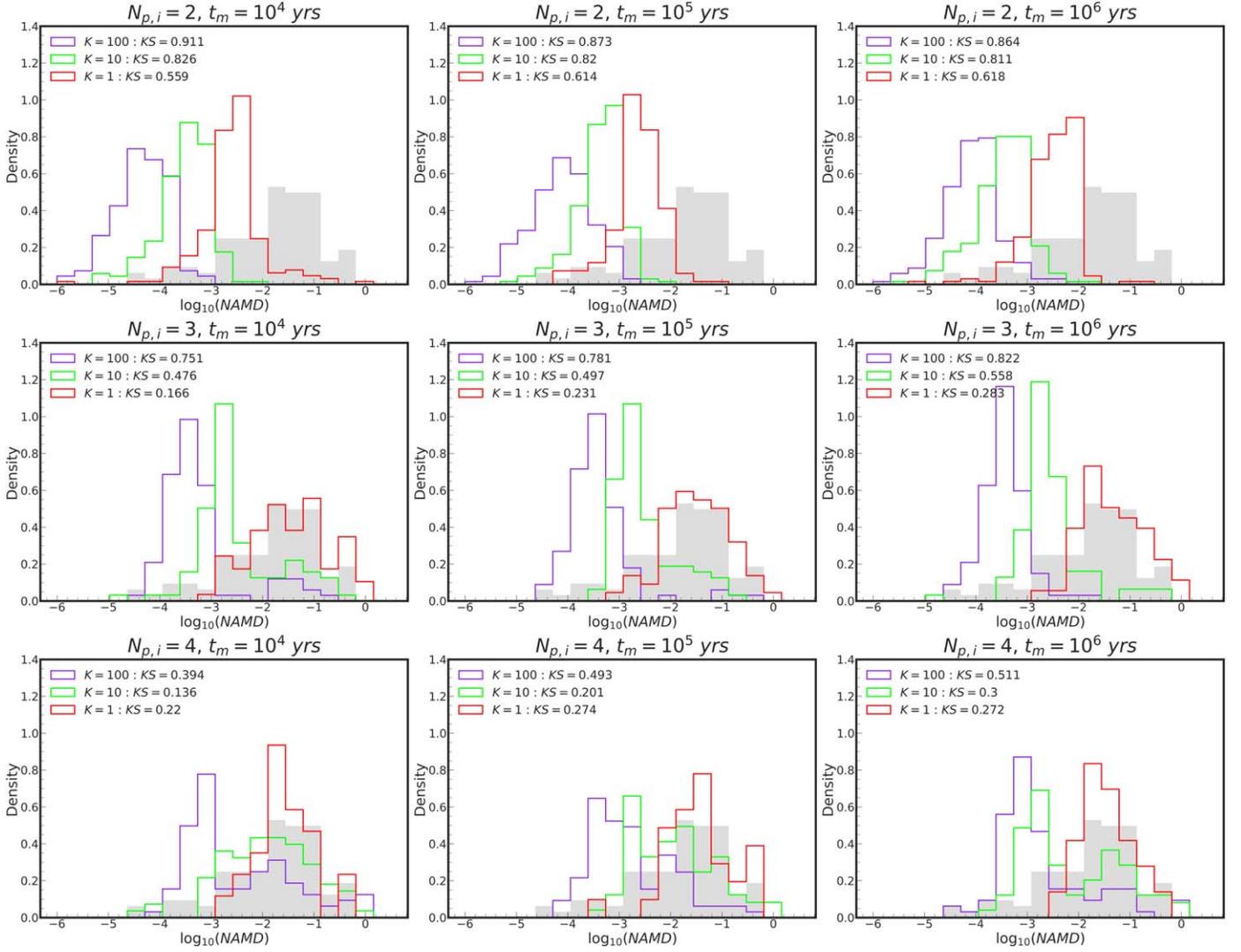

**Figure 6.** Normalized angular momentum deficit (NAMD) distributions for synthetic systems. The nine panels show distributions for subsamples obtained by selecting systems with the corresponding combinations of $t_m$ (migration time) and $N_{p,i}$ (initial multiplicity) from our simulations. In gray, on each panel, we plot the NAMD distribution for giant planets in the CLS sample. To aid comparison between observation and our simulations, we also report the value of the KS statistic for each subsample's NAMD distribution with respect to that of the CLS sample on the corresponding panel.

Upon examination, we find that, in some similarity to the previous section, combinations of high $N_{p,i}$, faster migration ($t_m = 10^4$ or $10^5$ yr), and weak damping ($K = 1$ or 10) are preferred in order to match the CLS giant planet NAMD distribution. The closest match between our simulations and observations (as measured by the KS statistic) is given by the N4T4K10 subsample, followed by N4T5K10 and N3T4K1, which present similarly good fits. This demonstrates how $N_{p,i}$, $t_m$, and $K$ can all contribute to determining the degree of excitation a system is likely to experience over its lifetime. Interestingly, subsamples with two-planet initial conditions (top row of Figure 6) broadly struggle to generate enough excitation to fit the observational NAMD distribution, a finding in line with those from Section 3. As a metric that jointly measures both period ratio and NAMD distribution similarity, we define $\mathcal{S} = \sqrt{\mathrm{KS}_{\mathrm{NAMD}}^2 + \mathrm{KS}_{P_{i+1}/P_i}^2}$ and report its value for every subsample in our simulations in Table 1. We find that N3T4K1 best minimizes $\mathcal{S}$, followed by N4T6K1, N3T5K1, N4T4K10, and N4T5K1.

---
[8] Though the systems in our simulations are purely coplanar, the observed giant planet systems do not have strong constraints on inclination. We thus make the assumption of coplanarity for the observational sample as well, cautioning that it may not necessarily hold.

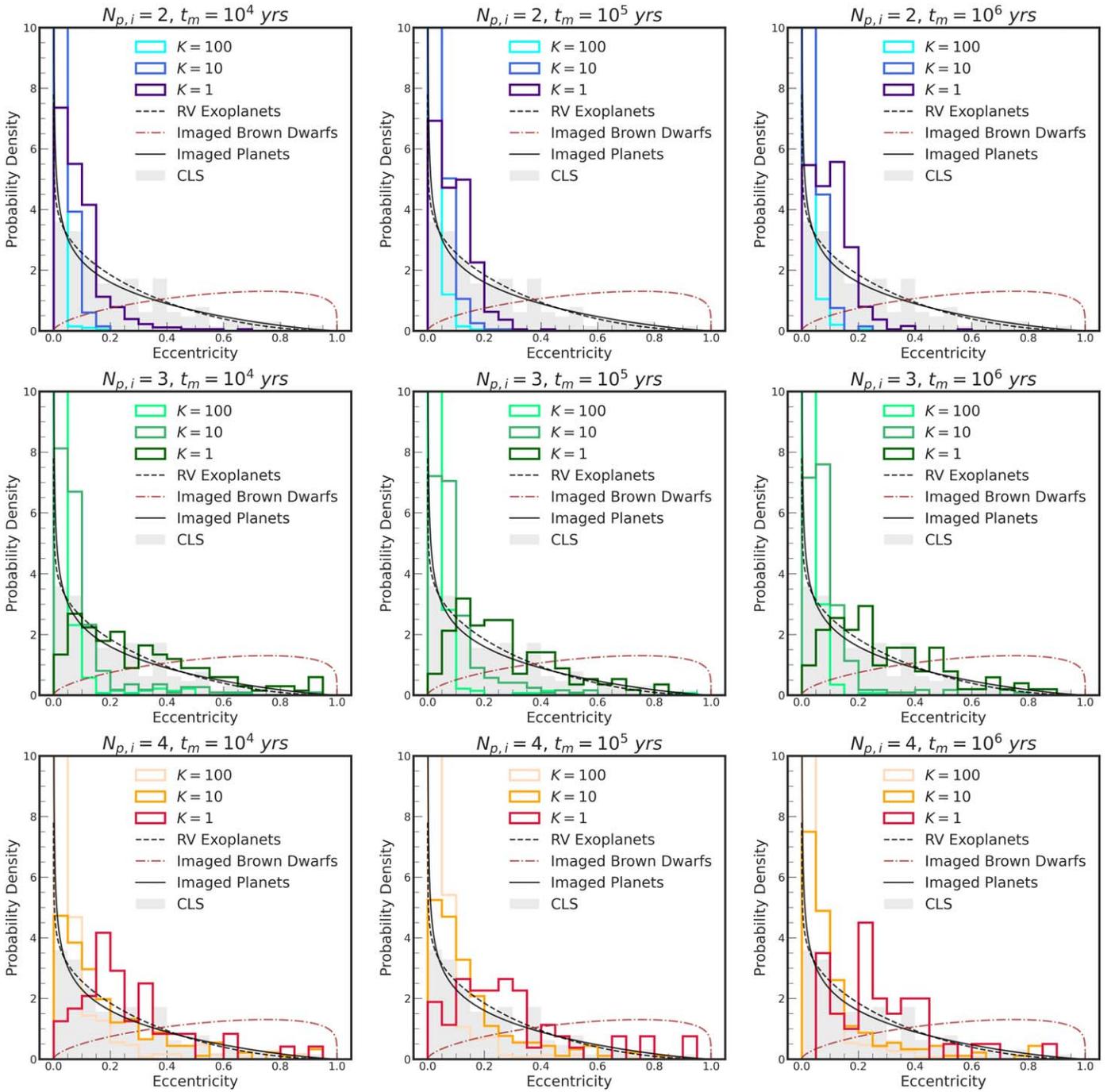

**Figure 7.** Final eccentricity distributions for our simulated systems. On each panel, we also show the eccentricity distribution of giant planets in the CLS sample, as well as the inferred eccentricity distributions for exoplanets characterized through the radial-velocity method (Kipping 2013), imaged planets (Nagpal et al. 2023), and imaged brown dwarfs (Bowler et al. 2020) for reference. We observe that two-planet initial conditions evolve into a population of planets that is systematically less eccentric than the eccentricity distributions inferred from exoplanet observations. In contrast, weakly damped high-multiplicity subsamples produce eccentricity distributions that qualitatively resemble the observed distributions.

These trends are echoed by the eccentricity distributions (Figure 7) for our synthetic systems. Two-planet initial conditions produce systems that are systematically less eccentric than both the CLS sample and previously inferred eccentricity distributions for exoplanets characterized using the radial velocity method (Kipping 2013) and direct imaging (Bowler et al. 2020; Do Ó et al. 2023; Nagpal et al. 2023).

Once again, high-multiplicity weakly damped initial conditions perform better: the eccentricity distributions for these subsamples prove a noticeably better match to the CLS and direct imaging samples.

## 5. Hot Jupiter Production

One proposed pathway for Hot Jupiter (HJ) formation is high eccentricity migration, in which planet–planet scattering perturbs the proto-HJ onto a highly elliptical orbit with a small enough periapsis to trigger the gradual tidal circularization of the orbit through planet–star interactions. Yet, whether dynamical instabilities in coplanar giant planet resonant chains can produce such proto-HJs at the requisite rate remains an open question. In an attempt to estimate the efficiency of this process, for each system in our simulations, we calculate the minimum periapsis distance $a(1-e)$ reached by any of its planets across their disk dispersal and gas-free evolution. We adopt $a < 0.1$ au as our definition of an HJ, which corresponds to an initial periapsis of $a(1-e) = 0.05$ au, as $a_{\rm final} \simeq 2a(1-e)$ for a planet undergoing tidal circularization (Dawson & Johnson 2018). Since all our simulated systems begin their gas-free evolution with the inner planet at 1 au, simply flagging planets that satisfy (at the end of the simulation) $a(1-e) < 0.05$ as HJ "progenitors" is certain to bias our estimates, as the inner edges of real giant planet systems do not exhibit such uniformity. In an attempt to address this bias, we employ the following empirically motivated scheme as part of our calculation:

1. For a given simulated system, randomly draw a scaling factor $a_s$ from an empirical CDF fit to the distribution of innermost semimajor axes[9] for the giant planet systems in the CLS sample.
2. Scale the semimajor axes of each planet in the system by $a_s$.
3. If, among the set of planets surviving the entire simulation for a given system, any satisfy $R_{\rm Roche} < a(1-e) < 0.05$ au at any point within the long-term gas-free integration, the system is flagged as an HJ progenitor.[10] In practice, however, we observe that virtually none of our surviving planets get scattered to periapses within the Roche lobe.

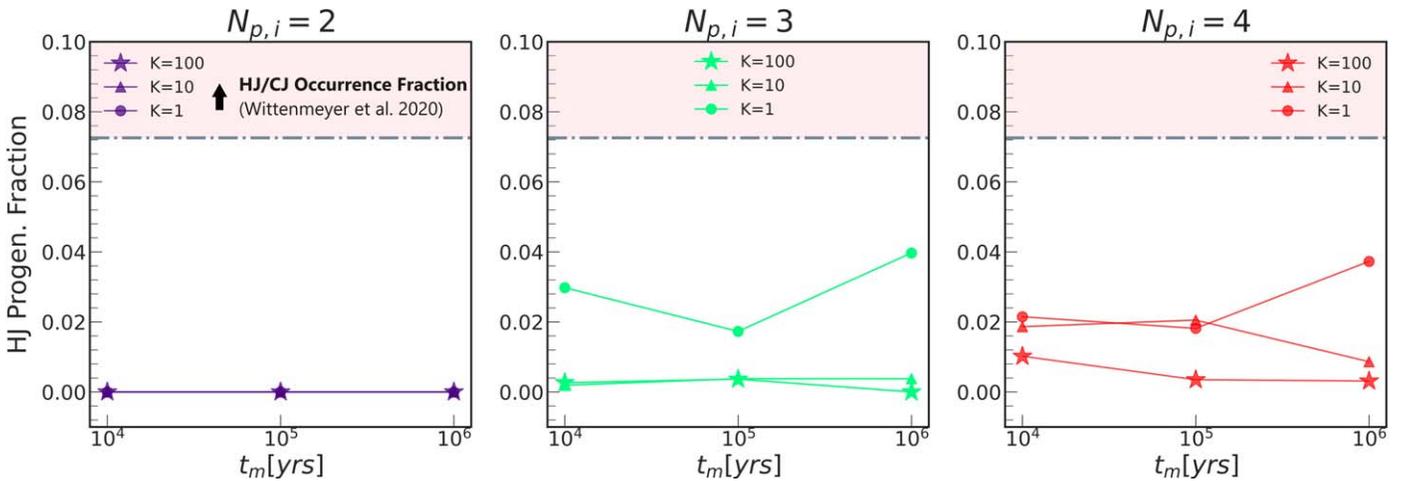

**Figure 8.** Estimates of Hot Jupiter production from our simulations, split up by: initial multiplicity ($N_{p,i}$; left to right), migration timescale ($t_m$; x-axis), and damping strength ($K$; marker). For comparison, we shade in pink the ratio between the observed occurrence rates of Cold Jupiters and Hot Jupiters from Wittenmyer et al. (2020).

We repeat the above procedure 1000 times for each subsample to estimate the HJ production rate; the results are summarized in Figure 8 and Table 1. We immediately observe that our synthetic systems struggle to produce HJs at rates commensurate with the ratio of Hot and Cold Jupiter occurrence rates calculated by Wittenmyer et al. (2020), which is shaded in pink in Figure 8. Even N3T6K1, the most efficient subsample, produces proto-HJs in only $4.0\% \pm 2.5\%$ of systems. This implies that dynamical instabilities within coplanar giant planet resonant chains may not produce enough Hot Jupiters to account for the measured relative occurrence of $\sim 8\%$ between Hot and Cold Jupiters (Wittenmyer et al. 2020). While planets in our simulations routinely reach small periapses during the scattering phase, they are almost always ejected before the end of the simulation and are thus not classified as HJ progenitors.

## 6. Discussion

In this exploration, we carried out dynamical simulations for a large suite of synthetic multigiant planet systems, integrating each system from end-to-end through phases of migration, disk dispersal, and long-term gas-free evolution. Within this framework, we studied the impact of varying $t_m$, $N_{p,i}$, and $K$ on the nature and fates of giant planet resonant chains formed through convergent migration. By comparing the final architectures of our simulated systems with properties of giant planets from the catalog of Rosenthal et al. (2021), we uncovered a marked preference for many-giant ($N_{p,i} \geqslant 3$) initial conditions coupled with weak damping on the order of $K \sim 1$–10.

In the last part of this study, we estimated the efficiency of coplanar planet–planet scattering at producing HJs from instabilities in resonant chains. Broadly speaking, our synthetic systems produce HJs at rates too low to account for the entirety of their observed occurrence. Therefore, our results disfavor (as a dominant pathway for HJ formation) high eccentricity migration induced by *coplanar* planet–planet interactions within giant planet resonant chains. Our finding is especially intriguing given the steady accumulation of studies (e.g., Rice et al. 2022; Zink & Howard 2023) that have found evidence to suggest that the observed population of HJs may have dominantly formed through high-eccentricity migration (Petrovich 2015). This discrepancy may indicate the need for large early-stage planetary mutual inclinations (as may emerge in systems with tilted inner disks; Benisty et al. 2023;

---
[9] Before drawing from the CLS distribution, we impose a cut at an orbital period of 30 days, in order to avoid contamination by systems in the sample that host HJs.
[10] We assume a planetary radius of $1 R_{\rm Jup}$ for this calculation, although all the planets in our simulations are point particles. We note that giant planet radii can exceed this value early in their lifetime by a factor of $\leqslant 2$.

Villenave et al. 2024; Zanazzi & Chiang 2024) or wide binary companions to trigger the additional dynamical excitation required to produce more HJs. Our results also leave open the possibility that spin–orbit misalignments may be primordial in nature (e.g., Lai et al. 2011; Batygin & Adams 2013; Spalding & Batygin 2015), an interpretation compatible with the analysis of Morgan et al. (2024).

Care must be taken in interpreting the results of this work. By virtue of attempting to place model-independent constraints on these quantities through the kind of simplified parametric framework used in this study, we forego modeling the effects of physical phenomena, which are likely to play a role in shaping the early stages of giant planet evolution. For example, Murray et al. (2022) found that disk precession could significantly alter resonance capture from the gas-free case and lead to larger amplitude libration—potentially paving the path toward easier instability triggering relative to the scenarios considered in this work. Additionally, chains of giant planets are likely to carve *mutual gaps* in a disk, the effect of which on planet-to-planet variations in the strength of eccentricity damping is still unclear, particularly in low-viscosity disks (Lega et al. 2021; Griveaud et al. 2023). Furthermore, recent detailed studies of the migration of super-Jupiter mass planets have found that such planets may migrate *outwards*. It is also possible that resonant chains formed by convergent migration do not always last for the entire lifetime of the disk: overstable librations (Deck & Batygin 2015) and changes in the migration rate (Kanagawa & Szuszkiewicz 2020) can both act to break resonances at an early stage. All these topics warrant further study. With all these caveats in mind, we shall now attempt to outline some of the implications of taking our results at face value.

If, as the results of our study imply, high multiplicities are the norm for young giant planet-hosting systems, a natural question arises: why do systems with multiple detected giant planets form such a small fraction of the observational sample? One possible resolution to this discrepancy may come from ejections: as Section 3 shows, the majority of our synthetic systems with $N_{p,i} = 4$ experience the ejection of two or more planets over the course of their monitored dynamical evolution. Interestingly, the recent work of Miret-Roig et al. (2022) and Pearson & McCaughrean (2023) uncovered a surprisingly large population of free-floating planets, a finding which suggests that ejections might be extremely common in the early evolution of planetary systems. Viewed in the light of the ubiquity of substructures in disks studied by ALMA, our requirements for high early multiplicities are especially interesting, given that the question of whether all such features can be linked to the presence of planets remains an active area of research.

Moving on to $t_m$, we find a weak overall preference for short convergence times on the order of $t_m = 10^{4\sim 5}$ yr. This is puzzling, considering that Type-II migration is thought to occur on significantly slower timescales (Ida et al. 2018; Kanagawa et al. 2018). Our recovered preference for short $t_m$ indicates that *tightly packed* initial resonant configurations are required for the onset of dynamical instabilities. While rapid migration timescales are needed for such configurations to emerge from the convergent migration prescription used for our simulations, hydrodynamical effects in real protoplanetary disks may disfavor capture into widely spaced higher-order resonances without the need for rapid migration. Keeping this possibility in mind, it is nevertheless curious that migration on timescales as rapid as $t_m = 10^{4\sim 5}$ yr is a consistent feature of hydrodynamic simulations modeling giant planet formation via the gravitational instability mechanism (Baruteau et al. 2011; Stamatellos 2015; Rowther & Meru 2020). Though this pathway has in recent years attracted criticism, recent advances on the theoretical front (e.g., Boss & Kanodia 2023; Longarini et al. 2023) together with intriguing discoveries from observations (e.g., Morales et al. 2019; Weber et al. 2023) are an indication that it deserves further study. We also observed that high initial giant multiplicities can help offset the need for rapid migration, as evidenced by the strong match between the N4T6K1 simulations, which had *slow* migration with $t_m = 10^6$ yr, and the CLS subsample.

Finally, the strong preference we unearth for simulations with $K \sim 1$ points toward scenarios in which giant planets in a disk rarely experience strong eccentricity damping. This finding is in line with those of Bitsch et al. (2020), who find a similar trend when simulating the evolution of giant planet systems within a pebble and gas accretion framework. Nevertheless, the full details of eccentricity damping for giant planets cannot be parameterized by a single number, and hydrodynamical simulations have demonstrated complex dependence on disk structure and accretion rate. As the galactic planetary census continues to come into sharper focus, it is increasingly clear that an integrated approach, combining observational insights with numerical modeling of planetary dynamics, is essential for a more comprehensive understanding of how planetary systems form. The results presented in this work constitute one step in this direction.

## Acknowledgments

V.N. would like to acknowledge Caltech SURF for supporting this research. We also thank the anonymous reviewer for the helpful comments, as well as Mario Flock, Sarah Millholland, Malena Rice, Sarah Blunt, B.J. Fulton, Lee Rosenthal, and Andrew Howard for helpful conversations. V.N. would also like to thank his fellow residents at International House for their great support, as well as for their commendable patience in listening to numerous impassioned monologues about resonant chains. K.B. is grateful to Caltech, the David and Lucile Packard Foundation, and National Science Foundation (grant number: AST 2109276) for their generous support. This research was enabled by the following software: rebound (Rein & Liu 2012), reboundX (Tamayo et al. 2020), numpy (Harris et al. 2020), pandas (McKinney 2010), matplotlib (Hunter 2007), and scipy (Virtanen et al. 2020). For the purpose of open access, the author has applied a Creative Commons Attribution (CC BY) license to any Author Accepted Manuscript version arising from this submission.

## ORCID iDs

Vighnesh Nagpal https://orcid.org/0000-0001-5909-4433
Max Goldberg https://orcid.org/0000-0003-3868-3663
Konstantin Batygin https://orcid.org/0000-0002-7094-7908